\documentclass[twocolumn,twoside]{revtex4}
\usepackage{graphicx}
\usepackage{fancyhdr}
\usepackage{amsmath}

\begin{document}

\title{Pentaquarks}

%

\author{Lorenzo Capriotti\\
on behalf of the LHCb Collaboration\\
with results from the Belle Collaboration}
\affiliation{Alma Mater Studiorum Universit\`a di Bologna, Dipartimento di Fisica e Astronomia, via Irnerio 46, Bologna, Italy\\ INFN - Sezione di Bologna, viale B. Pichat 6/2, Bologna, Italy}

\begin{abstract}
 
\end{abstract}

\maketitle

\thispagestyle{fancy}


\section{Introduction}
\subsection{Exotic charm spectroscopy}
\begin{figure}
\centering
\includegraphics[width=80mm]{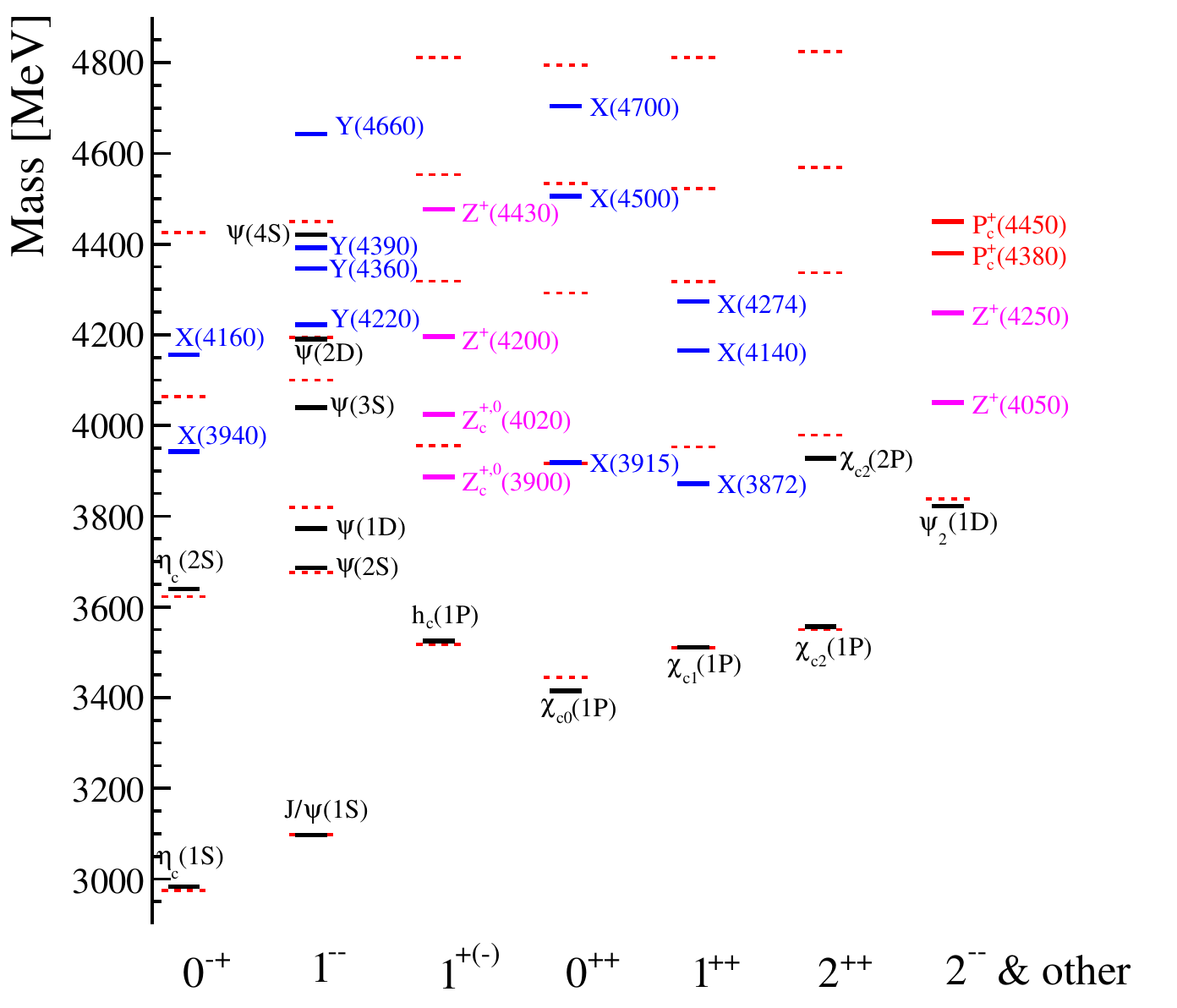}
\caption{The charmonium spectrum (2017). Black lines represent observed $c\bar{c}$ states, blue (magenta) lines are observed neutral (charged) exotics and the red dotted lines represent predictions according to Eq.~\ref{eq:Cornell}. Two observed pentaquark candidates are shown in red. Adapted from~\cite{spectrum}}
\label{fig:ccspectrum}
\end{figure}

The excitation spectrum of a charm-anticharm state is well described by a semi-relativistic phenomenological potential (Cornell potential)~\cite{cornell}, developed in the 70's after the discovery of the $J/\psi$ and other charmonia states. It comprises a sum of three terms:
\begin{equation}
V(r) = -\frac{4}{3} \frac{\alpha_s(r)}{r} + \sigma r + \delta(1/r^2),
\label{eq:Cornell}
\end{equation}
where the first is a short-distance colour potential, the second is a long-distance confinement term and the third includes spin-spin and spin-orbit corrections. This potential is found to be particularly accurate to describe the spectrum of excited $c\bar{c}$ and $b\bar{b}$ states, as well as predict new states before their observation. However, in the last two decades a large number of states has been discovered, which are compatible with being composed by - at least- a $c\bar{c}$ pair as constituent quarks as they all decay into a final state with a charmonium; however, they do not fit in the expected spectrum given their mass and other properties, such as decay and production rates or quantum numbers, which would be expected from a pure $c\bar{c}$ state. All these states are labelled as \emph{exotic states}. Fig.~\ref{fig:ccspectrum} shows the charmonium spectrum, including all exotic states, as it was in 2017.
\subsection{Multiquark candidates}
The first exotic state ever observed is the $X(3872)$ (recently renamed $\chi_{c1}(3872)$), from data analysed by the Belle experiment, in 2003~\cite{belleX}. Its mass, $3871.69$ MeV/$c^2$, is not close to any predicted $c\bar{c}$ state. Its width is extremely narrow, and indeed it is limited by experimental resolution to be less than 1.2 MeV/$c^2$; this is unexpected for a $c\bar{c}$ state above the open charm mass threshold. Moreover, the measured radiative decay rates seem not match predictions~\cite{Xrad}, and it has been observed to decay into two final states with different isospin ($J/\psi \rho^0$ and $J/\psi \omega$) with similar rates. Given its proximity to the $D^0 \bar{D}^{*0}$ mass threshold, it is believed that the $X(3872)$ is a good candidate for a bound state of four quarks (tetraquark), although the binding mechanism is still unknown (see next section). Many of the observed exotic states are tetraquark candidates, and in particular the $Z_c$ states (for instance, $Z_c(4430)$) which are charged and must have a minimal constituent quarks content of $c\bar{c}d\bar{u}$. Similarly, the charged states discovered by the LHCb experiment in the $J/\psi p$ final state must have a minimal quark content of $uudc\bar{c}$; therefore, they are pentaquark candidates. 

More information on exotic states, along with details about the models described in the next section, can be found in Ref.~\cite{XYZ}.
\subsection{Models for multiquark states}
Several models have been proposed to describe the multiquark candidates. The two main interpretations are described in this section as an example.

In the \emph{compact multiquark} picture, the quarks are tightly bound into a colour-neutral state. Large widths are expected, as well as many isomultiplet states with similar mass. While this model can explain the large prompt production of some candidates, it fails to describe the narrow width observed in many exotic states; furthermore, no isospin partner has ever been found.

In the \emph{molecular} picture, the exotic state is described by a mesonic - or baryonic - molecule. The bound state is expected to have mass a little below the mass threshold necessary to decay strongly into its components; therefore, at least one couple of quarks must change their confining partners before the molecule can decay, and this leads to long-lived states (i.e. narrow widths). Few states are predicted, due to the low binding energy required to form them, which allows the existence of only low orbital momentum states, perhaps only S-wave. It is believed that the binding energy is too small to account for the large prompt production in colliders. Other models are in principle allowed, as well as mixture of different models.

\section{First observation of pentaquarks in $\Lambda_b^0 \to J/\psi K^- p$ decays}
\label{sec:penta1}
\subsection{Direct observation}
\label{sec:direct}
\begin{figure}
\centering
\includegraphics[width=80mm]{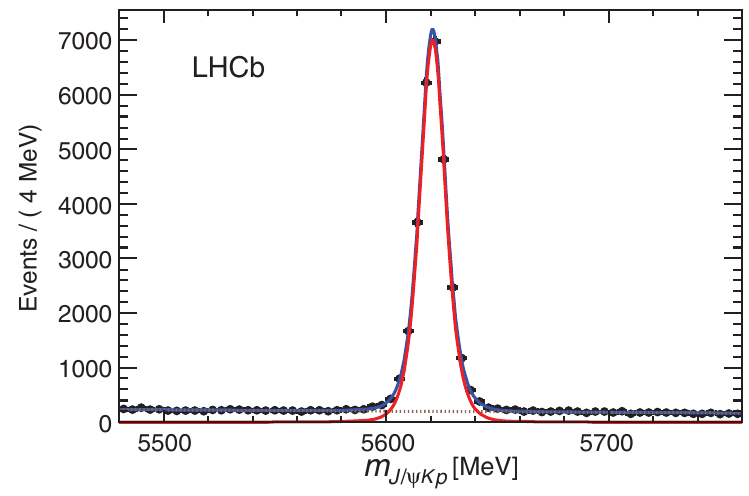}
\caption{Fit to the $\Lambda_b^0$ peak, used to extract signal weights.}
\label{fig:lambdab}
\end{figure}

The LHCb experiment has analysed $\Lambda_b^0 \to J/\psi K^- p$ decays from $pp$ collisions using the full Run 1 dataset, which comprises 3 fb$^{-1}$ of data collected at 7 and 8 TeV during 2011 and 2012~\cite{penta1}. This channel is expected to be dominated by resonant $\Lambda^* \to K^-p$ decays; however, a large structure is visible also in the $m_{J/\psi p}$ spectrum. In order to investigate the nature of said structure, it is necessary to chech whether this can be a result of reflections in the Dalitz plot caused by the expected $\Lambda^*$ activity. In order to do so, 14 well established $\Lambda^*$ are taken into account in building a six-dimensional amplitude model, which includes 5 decays angles in the helicity formalism~\cite{helicity} and $m_{Kp}$. The amplitude fit is performed following two strategies: first, a sum of a double-sided Hypatia function~\cite{hypatia} and a background component, modeled on sidebands, is fit to the $\Lambda_b^0$ peak. From this fit, signal weights are extracted with the $sPlot$ method~\cite{splot}. The full amplitude model is then fit to background-subtracted data, obtained by applying the aforementioned weights. The fit to the $\Lambda_b^0$ peak is shown in Fig.~\ref{fig:lambdab}. For the second strategy, the amplitude model is fit to unweighted data, with an additional background component, within $\pm 2 \sigma$ of the $\Lambda_b^0$ mass peak. The two strategies give compatible results.

\begin{figure}
\centering
\includegraphics[width=80mm]{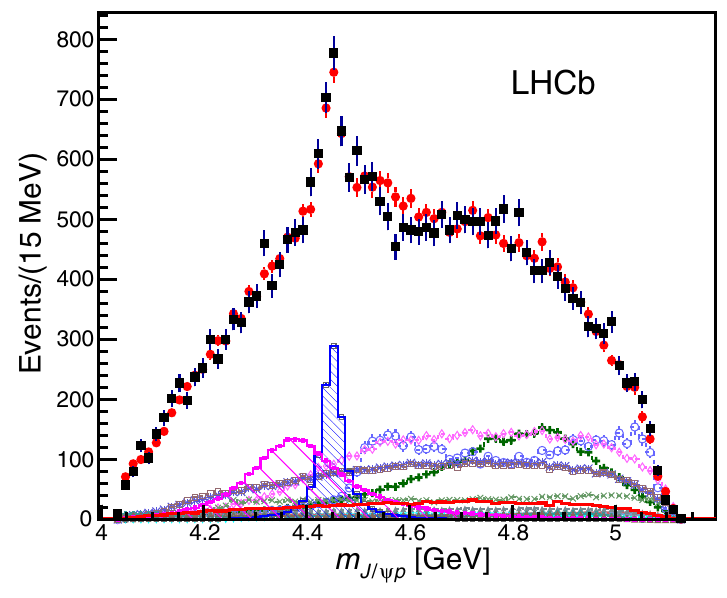}
\caption{Distribution of $m_{J/\psi p}$ (black) and fit projection (red). The two shaded histograms represent the pentaquarks contributions.}
\label{fig:ampfit}
\end{figure}

By studying the $m_{J/\psi p}$ projection of the six-dimensional fit, shown in Fig.~\ref{fig:ampfit}, it is observed that the fit does not describe the data well unless two more contributions are added: a broad state, $\Gamma=205 \pm 18$ MeV, with mass 4380 MeV/$c^2$, and a narrow state, $\Gamma=39 \pm 5$ MeV, with mass 4450 MeV/$c^2$. These two states are labelled, respectively, $P_c(4380)^+$ and $P_c(4450)^+$. Their favourite $J^{PC}$ assignment is $3/2^-$ and $5/2^+$, though combinations ($3/2^+$, $5/2^-$) and ($5/2^+$, $3/2^-$) are also possible. The singificance of each contribution is, respectively, 9$\sigma$ and 12$\sigma$.
\subsection{Model-independent confirmation}
The analysis described in Section~\ref{sec:direct} is repeated using a different, model-independent approach~\cite{independent}. This consists in describing the bidimensional plane ($m_{Kp}$,$\cos\theta_{\Lambda^*}$) by expanding the helicity angle $\theta_{\Lambda^*}$ in Legendre polynomials,
\begin{equation}
\frac{dN}{d ( \cos\theta_{\Lambda^*})} = \sum_{l=0}^{l_{max}} a_l P_l(\cos\theta_{\Lambda^*}),
\label{eq:LegPol}
\end{equation}
where $P_l$ is the Legendre polynomial of order $l$ and the coefficients of the expansion (Legendre moments) are defined as
\begin{equation}
a_l = \frac{2l+1}{2} \int_{-1}^{+1} \frac{dN}{d ( \cos\theta_{\Lambda^*})} P_l(\cos\theta_{\Lambda^*}) \text{ }d\cos\theta_{\Lambda^*}.
\label{eq:LegCoef}
\end{equation}

\begin{figure}
\centering
\includegraphics[width=80mm]{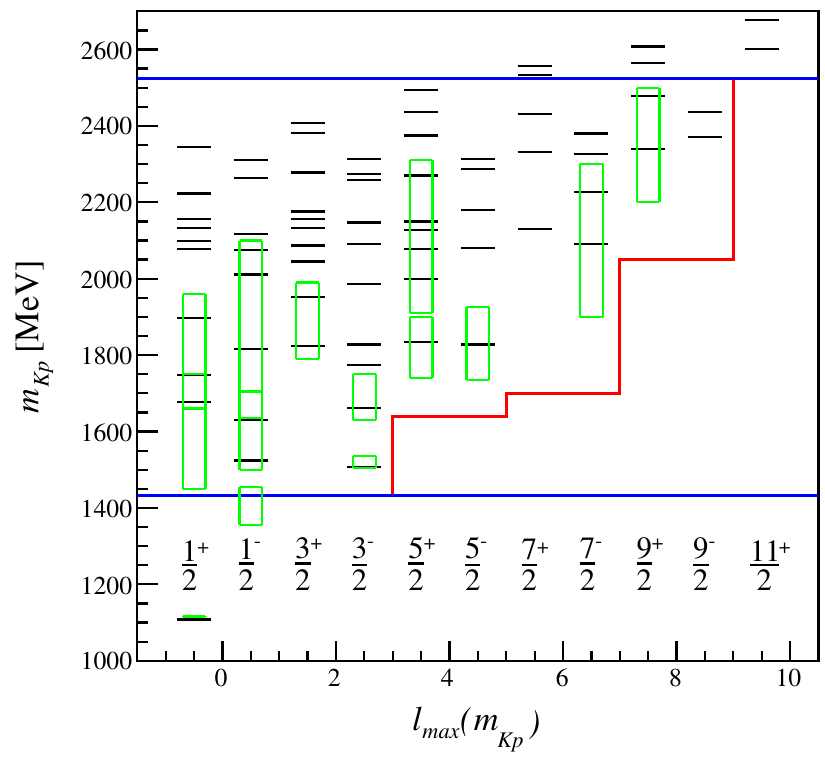}
\caption{Definition of $l_{max}$, used in the Legendre expansion, as function of both predicted (black) and well-established (green) $\Lambda^*$ resonances available for a certain value of $m_{K^-p}$. The blue lines define the mass range. The red line shows the $l_{max}$ filter.}
\label{fig:jmax}
\end{figure}

The moments $a_l$ are extracted from efficiency-corrected, background-subtracted data by quadratically interpolate the $m_{Kp}$ histogram. Under the hypothesis that all $\Lambda_b \to J/\psi p K^-$ decays proceeds via intermediate $\Lambda^+$, $\Sigma^*$ or non resonant $K^-p$, each component cannot contribute to moments of rank higher than $2J_{max}$, where $J_{max}$ is the highest spin of any $K^-p$ contribution at a given $m_{Kp}$ value; therefore, $l_{max}$ in Eq.~\ref{eq:LegPol} depends on the spin of the intermediate resonances that might be available at a certain value of the $K^-p$ mass. This is summarised in Fig.~\ref{fig:jmax}.

\begin{figure}
\centering
\includegraphics[width=80mm]{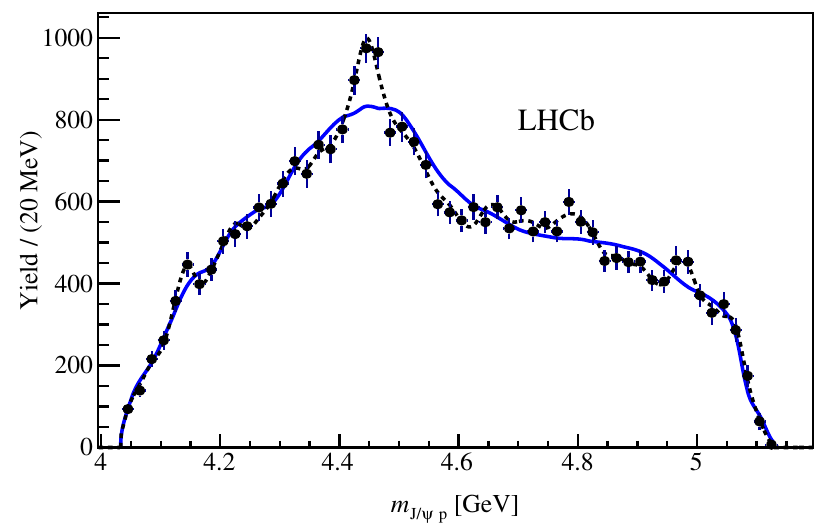}
\caption{Efficiency-corrected and background-subtracted $m_{J/\psi p}$ distribution, with the Legendre expansion (blue) and the second unphysical expansion with $l_{max}=31$ (black dotted).}
\label{fig:modind}
\end{figure}

If exotic contributions are present, then the hypothesis under which the calculation of the Legendre moments has been performed will not be valid; therefore, the $m_{J/\psi p}$ distribution will not be well described by the expansion. This is indeed what is observed, as can be seen in Fig.~\ref{fig:modind}: the discrepance is calculated using a second Legendre expansion with an unphysically large $l_{max}=31$ which describes the statistical significant features of data; its significance is calculated to be $9\sigma$, thus confirming the presence of contributions to $\Lambda_b \to J/\psi p K^-$ decays, either due to exotic resonances or rescattering effects. The latter are investigated and excluded in separate results, described in the next Section.
\section{Rescattering effects}
The narrow structure observed by LHCb with mass 4450 MeV/$c^2$ happens to be located exactly at the $\chi_{c1}p$ mass threshold. This can be a signal of a kinematic enhancement due to rescattering effects~\cite{resc1,resc2}, which can happen if the $\Lambda_b^0$ decays to $\chi_{c1}\Lambda^*$, and then the proton from $\Lambda^* \to K^-p$ scatters with the $\chi_{c1}$ via photon exchange, forming the $J/\psi p$ final state. In order to have a threshold enhancement all intermediate particles ($\Lambda^*,p,\chi_{c1}$) must be on shell; furthermore, the $\Lambda^*$ mass must lie within a kinematically allowed range and amongst the excited $\Lambda$ states, one happens to satisfy this requirement: $\Lambda(1890)$. The LHCb Collaboration has published two papers aiming at confirming or denying this effect regarding the pentaquark candidate $P_c^+(4450)$, and they are described below.
\subsection{Search for $P^+_c \to \chi_{c1}p$}
\begin{figure}
\centering
\includegraphics[width=80mm]{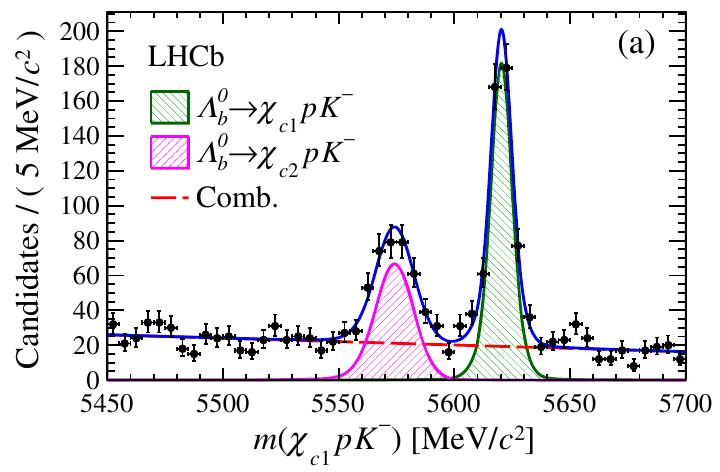}
\caption{The $m_{\chi_{c1}pK^-}$ distribution with overlaid fit (blue) and its components. Notice that, due to the $\chi_{c1}$ mass constraint, the $\Lambda_b^0$ peak is shifted towards low masses when decaying via a final state containing a $\chi_{c2}$.}
\label{fig:chi}
\end{figure}

If $P_c^+(4450)$ is a real resonance, then it could in principle decay to $\chi_{c1}p$, thus neglecting final-state interactions between $\chi_{c1}$ and $p$. The LHCb Collaboration has analysed the full Run 1 dataset searching for resonant structures in $\Lambda_b^0 \to \chi_{c1}pK^-$ decays~\cite{chic}. The number of observed signal decays, N$(\Lambda_b^0 \to \chi_{c1}pK^-) = 453 \pm 25$, is not large enough to allow an analysis of the $m_{\chi_{c1}p}$ spectrum; this will be updated with the addition of Run 2 data. The fit used to extract the signal yield is shown in Fig~\ref{fig:chi}. Nonetheless, this first investigation on this channel lead to the first observation of the decays $\Lambda_b^0 \to \chi_{c1,2}pK^-$ and first measurement of their relative branching fraction:
\begin{eqnarray}
\frac{\mathcal{B}(\Lambda_b^0 \to \chi_{c1} p K^-)}{\mathcal{B}(\Lambda_b^0 \to J/\psi p K^-)} = 0.242 \pm 0.014 \pm 0.013 \pm 0.009, \nonumber \\
\frac{\mathcal{B}(\Lambda_b^0 \to \chi_{c2} p K^-)}{\mathcal{B}(\Lambda_b^0 \to J/\psi p K^-)} = 0.248 \pm 0.020 \pm 0.014 \pm 0.009. \nonumber 
\end{eqnarray}
\subsection{Analysis of the Cabibbo-suppressed channel}
An observation of the $P_c(4450)^+ \to J/\psi p$ decay from $\Lambda_b^0 \to J/\psi p \pi^-$ decays would not only confirm the existence of an unexpected contribution, but also its exotic nature. In fact, while the Cabibbo-favoured channel is dominated by $\Lambda^* \to pK^-$ decays, the Cabibbo-suppressed channel is dominated by intermediate excited nucleons decays, $N^* \to p \pi^-$. The masses of the known excited nucleons all lie outside the range that would allow a kinematic enhancement due to rescattering; therefore, this mechanism would be much harder to accommodate in this picture.

\begin{figure}
\centering
\includegraphics[width=80mm]{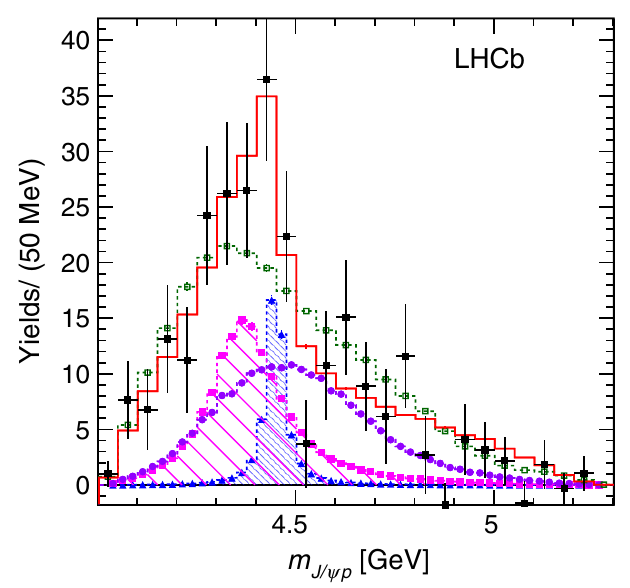}
\caption{Distribution of $m_{J/\psi p}$ for  and overlaid fits, with (red) and without (green dotted) the exotic contributions. The contribution from the two pentaquarks are shown as shaded, while the $Z_c(4200)^-$ as the purple histogram.}
\label{fig:CS}
\end{figure}

The analysis is performedy with the full LHCb Run~1 dataset~\cite{cabibbo}. The analysis strategy is similar to the one described in Section~\ref{sec:direct}: 14 well established $N^*$ resonances are used to build a six-dimensional amplitude model, using five decay angles in the helicity formalism and the $\pi^- p$ invariant mass; thanks to the $\Delta I = 1/2$ rule~\cite{deltaI}, the $\Lambda^* \to p \pi^-$ contributions are suppressed. This model is then fit to background-subtracted data, obtained by extracting signal weights with the $sPlot$ technique via a fit to the $\Lambda_b^0$ peak. Analogously to the Cabibbo-favoured case, the amplitude fit is found to be significantly improved when the two previously observer pentaquark candidates, $P_c(4380)^+$ and $P_c(4450)^+$, along with a $Z_c(4200)^- \to J/\psi \pi^-$ contribution, are added to the fits. The fit is shown in Fig.~\ref{fig:CS}, including the cut $m_{p\pi^-} > 1.8$ GeV/$c^2$ in order to enhance the exotic components. The combined significance obtained by adding the three exotic contributions is calculated to be 3.1$\sigma$. The production rates are found to be compatible with expectations based on the analysis of the Cabibbo-favoured channel.
\section{Recent searches for strange and beauty pentaquarks}
In principle, pentaquarks can exist also in the strange and the beauty sector. The most recent searches are presented in the following sections.  
\subsection{Search for $s$-flavoured pentaquarks}
\begin{figure}
\centering
\includegraphics[width=73mm]{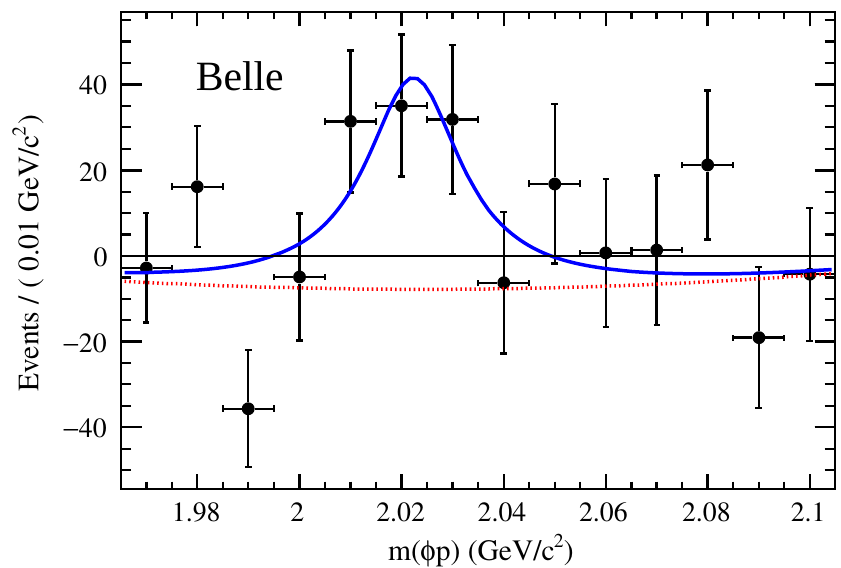}
\caption{Fit to the $m_{\phi p}$ spectrum for background-subtracted $\Lambda_c^+ \to \phi p \pi^0$ decays in a mass window of 20 MeV/$c^2$ around the $\phi$ peak. The total fit is shown in blue, while the phase space component in red.}
\label{fig:belle}
\end{figure}

The Belle Collaboration has analysed 915 fb$^{-1}$ of data collected at the $\Upsilon(4S)$ and $\Upsilon (5S)$ peaks to search for strange-flavoured pentaquarks in the strange-flavour analogue channel of the $P_c^+$ discovery, i.e. $\Lambda_c^+ \to \phi p \pi^0$~\cite{belle}. The hypothetic $P_s^+$ could be observed as a peak in the $m_{\phi p}$ spectrum, provided that the same production mechanism as the charm pentaquarks holds and if its mass is $m_{P_s^+} < m_{\Lambda_c^+} - m_{\pi^0}$. To extract the signal yield, a bidimensional fit to the variables $m_{K^+K^-p\pi^0}$ and $m_{K^+K^-}$ is performed. The number of signal events is found to be $148.4 \pm 61.8$. A sum of a relativistic Breit-Wigner and a phase space distribution determined from simulation is then fitted to the background-subtracted $m_{\phi p}$ spectrum, in a 20 MeV/$c^2$ mass window around the $\phi$ peak, in order to search for a putative $P_s^+$ peak. No clear evidence is found, and an upper limit at 90\% CL is set on the product of the branching fractions, normalised using $\Lambda_c^+ \to p K^- \pi^+$ decays,
\begin{equation*}
\mathcal{B}(\Lambda_c^+ \to P_s^+ \pi^0) \times \mathcal{B}(P_s^+ \to \phi p) < 8.3 \times 10^{-5}.
\end{equation*}
The fit to the $m_{\phi p}$ spectrum is shown in Fig.~\ref{fig:belle}.
As a reference, for the $P_c(4450)^+$ discovery the analogous quantity reads:
\begin{equation*}
\mathcal{B}(\Lambda_b^0 \to P_c^+ K^-) \times \mathcal{B}(P_c^+ \to J/\psi p) = (1.3 \pm 0.4) \times 10^{-5}.
\end{equation*}
\subsection{Search for $b$-flavoured pentaquarks}
According to the Skyrme model~\cite{skyrme}, the heavier the consituent quarks are, the more tightly bound the pentaquark state is; furthermore, no search for $b$-flavoured pentaquark exists in literature. The LHCb Collaboration has analysed the full Run 1 dataset to search for four possible, weakly-decaying $b$-flavoured pentaquarks~\cite{bpenta}. The weak decay channels, quark contents and mass search windows are summarised in Tab.~\ref{tab:bpenta}. The mass windows are chosen to be below the strong decays thresholds, and the subscript on each state indicates the final state the pentaquark would predominantly decay into if it had sufficient mass to decay strongly in those states.
\begin{table}
\caption{Quark content of the $b$-flavoured pentaquarks, weak decay modes and mass search windows (in MeV/$c^2$)}
\begin{tabular}{ccc}
\toprule
Quark content&Decay Mode&Search window\\
\hline
$\bar{b}duud$&$P_{B^0p}^+ \to J/\psi K^+ \pi^- p$&4668-6220\\
$b \bar{u} udd$&$P^-_{\Lambda_b^0 \pi^- } \to J/\psi K^- \pi^- p$&4668-5760\\
$b \bar{d} uud$&$P^+_{\Lambda_b^0 \pi^+} \to J/\psi K^- \pi^+ p$&4668-5760\\
$\bar{b}suud$&$P^+_{B^0_s p} \to J/\psi \phi p$&5055-6305\\
\hline
\end{tabular}
\label{tab:bpenta}
\end{table}
While there are many possible decay modes of these states, the analysis is focussed only on $b \to c\bar{c}s$ processes in order to have a $J/\psi$ in the final state, for which the LHCb experiment has large reconstruction efficiencies and reduced backgrounds. 

\begin{figure}
\centering
\includegraphics[width=80mm]{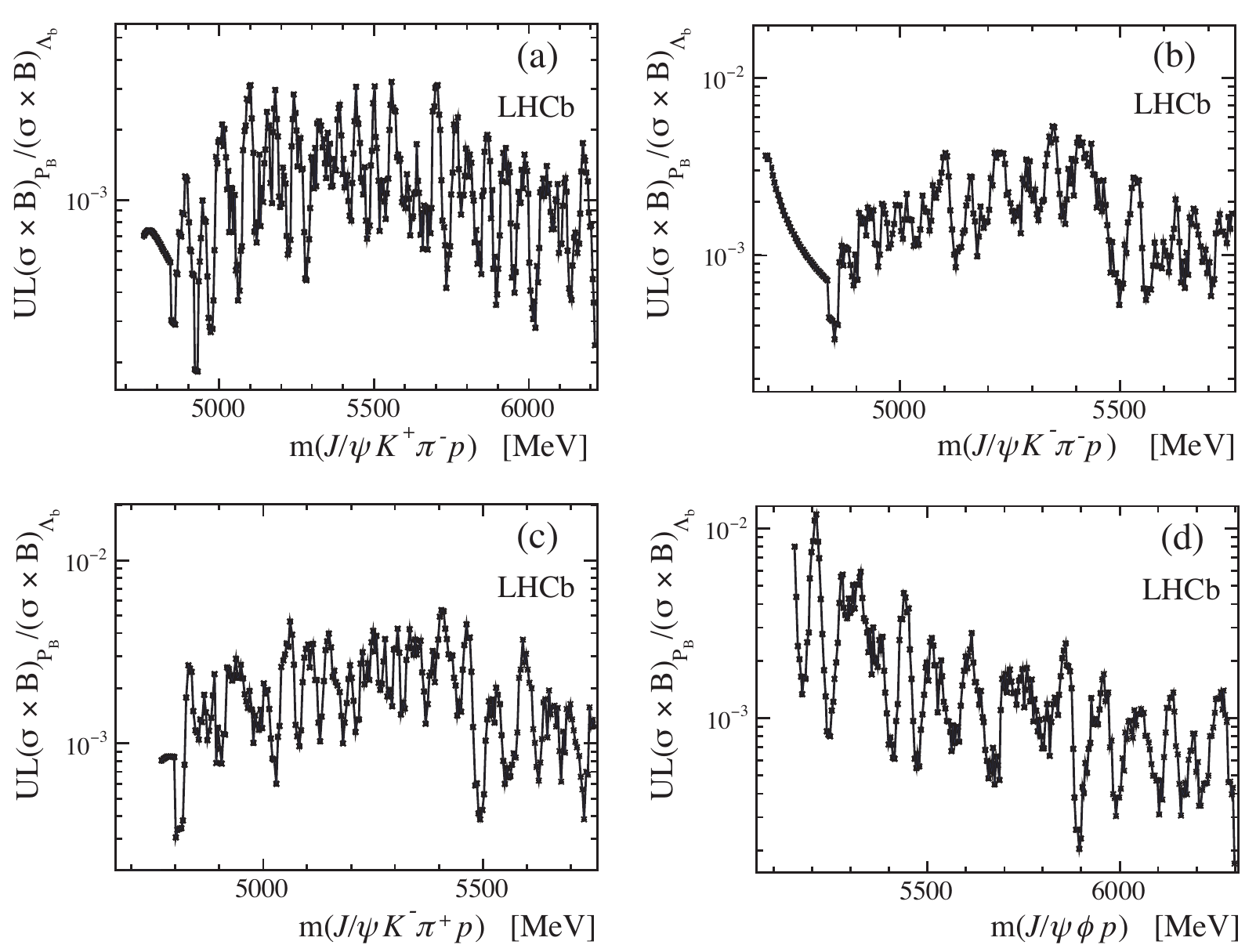}
\caption{Upper limits at 90\% CL on $\sigma \times \mathcal{B}$ for the four $b$-flavoured pentaquarks as a function of the invariant mass of their respective final state.}
\label{fig:bpentaUL}
\end{figure}

No signal is observed for any of the final states considered, and upper limits at 90\% CL are set on the products of the production cross sections and the branching ratios, normalised using $\Lambda_b^0 \to J/\psi p K^-$ decays. The upper limits as a function of the invariant masses in the search windows are shown in Fig.~\ref{fig:bpentaUL}.
\section{Observation of pentaquarks in $\Lambda_b^0 \to J/\psi K^- p$ decays (Run 1 and Run 2)}
The analysis presented in Section~\ref{sec:direct}, published in 2015, was updated including the full LHCb Run~2 dataset collected in the period 2015-2018 with a centre-of-mass energy of 13 TeV, resulting in a total integrated luminosity of 9 fb$^{-1}$~\cite{penta2}. The threefold increase in luminosity, an improved data selection and the increase in the production cross-section passing from 7-8 TeV to 13 TeV gives a ninefold increase in statistics. 
\begin{table}
\caption{Pentaquarks properties, from the fit in Fig.~\ref{tab:penta2}.}
\begin{tabular}{cccc}
\toprule
State&Mass [MeV/$c^2$]&Width [MeV/$c^2$]\\
\hline
$P_c(4312)^+$&$4311.9 \pm 0.7^{+6.8}_{-4.5}$& $9.8 \pm 2.7^{+3.7}_{-4.5}$\\
$P_c(4440)^+$&$4440.3 \pm 1.3^{+4.1}_{-4.7}$& $20.6 \pm 4.9^{+8.7}_{-10.1}$\\
$P_c(4457)^+$&$4457.3 \pm 0.6^{+4.1}_{-1.7}$& $6.4 \pm 2.0^{+5.7}_{-1.9}$\\
\hline
\end{tabular}
\label{tab:penta2}
\end{table}
The same amplitude model used in the 2015 analysis is fit to the new dataset, with the new data selection, and the results are found to be compatible; however, the increase in available statistics allows for new features to be observed in the $m_{J/\psi p}$ spectrum. A new narrow structure emerges at 4.3 GeV/$c^2$, and the former $P_c(4450)^+$ peak is resolved into two narrow peaks. 

The new structures are so narrow that they cannot be caused by artificial reflections in the Dalitz plot; for this reason, they are analysed with a one-dimensional fit rather than full amplitude fit. The latter is still necessary to measure the quantum numbers, and it will be done in a separate study given the computational difficulties linked with the level of precision required. 
\begin{figure}[!h]
\centering
\includegraphics[width=73mm]{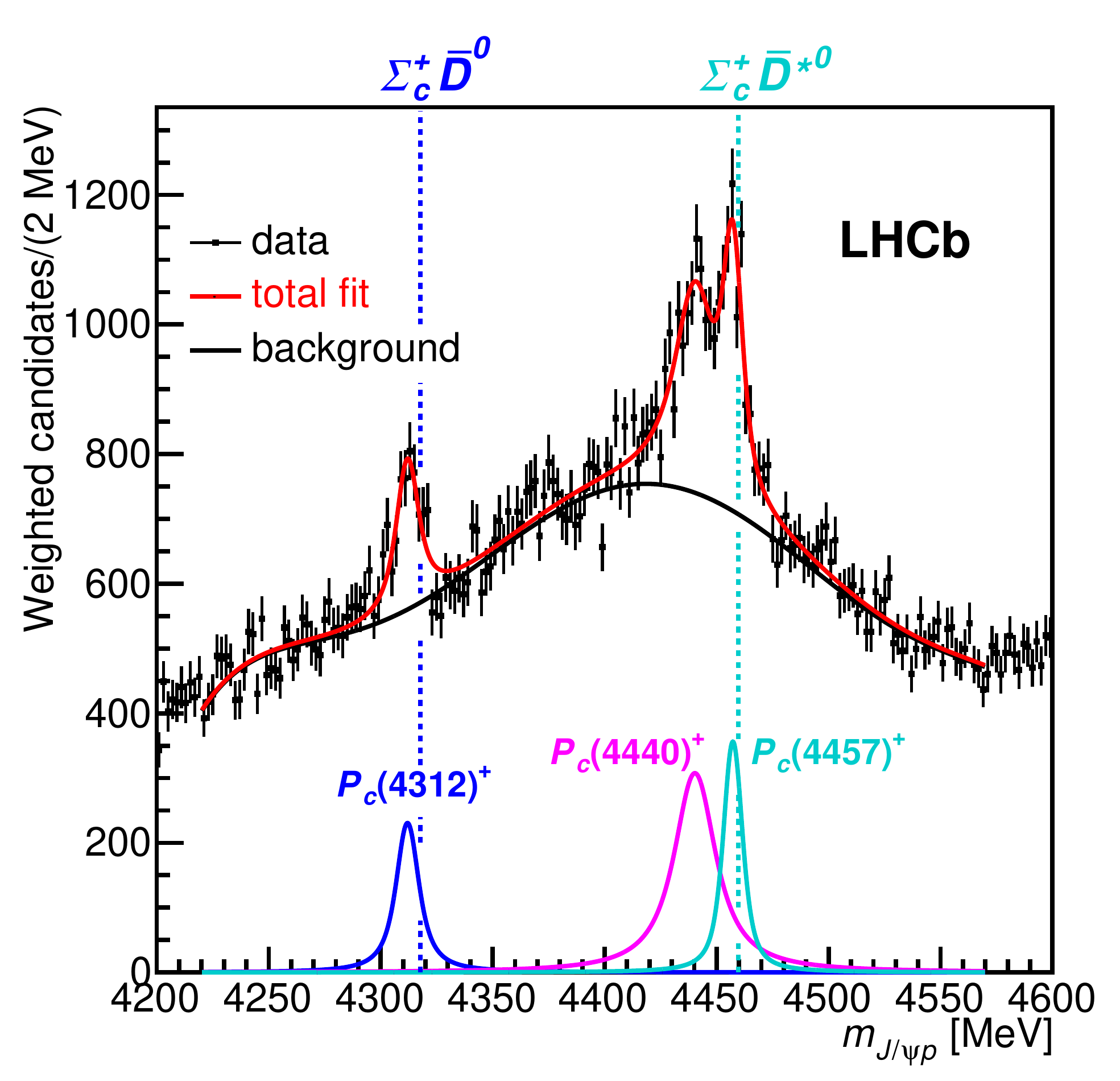}
\caption{Fit to the weighted $m_{J/\psi p}$ distribution. The vertical lines indicate the two mass thresholds.}
\label{fig:penta}
\end{figure}

The fit model is a sum of three Breit-Wigner functions and a sixth-order polynomial background. To enhance the signal sensitivity in data, weights are applied to the $\cos \theta_{P_c}$ (cosine of $P_c^+$ helicity angle) of each candidate: they are calculated as the inverse of the expected background at each value of $\cos \theta_{P_c}$, which is approximately the inverse of the candidates density as the signal represents just a few percent of the whole dataset. This is justified by noticing that candidates with $\Lambda^*$ will have predominantly positive $\cos \theta_{P_c}$. The fit results are shown in Table~\ref{tab:penta2} and in Fig.~\ref{fig:penta}. The latter shows also the mass thresholds of the $\Sigma_c^+\bar{D}^{(*)0}$ systems, and it can be noticed that the masses of the narrow peaks are just below these thresholds, which is what the molecular interpretation of these states would predict. 

\section{Conclusions}
The field of exotic spectroscopy is extremely rich and productive. Several observations and searches for pentaquark states have been performed in the last 4 years. This is still quite a recent discovery, and possibly the beginning of a new era in both observation of new states and understanding of the QCD binding mechanisms. 

Finally, regarding the interpretation of the pentaquark states, although their real nature is still unknown and the compact pentaquark model is still strictly not ruled out, the LHCb Run 2 update measurement provides the strongest evidence so far towards a molecular interpretation of the $P_c^+$ states, given by their narrow widths and masses close to the relative thresholds.

\bigskip 

\end{document}